\documentclass[aps,prL,twocolumn,superscriptaddress]{revtex4-1}
\usepackage{graphicx}
\usepackage{dcolumn}
\usepackage{bm}
\usepackage{amsmath}
\usepackage[english]{babel}
\usepackage[utf8]{inputenc}
\usepackage{verbatim}

\usepackage{layouts}

\begin{document}
	\bibliographystyle{achesmo}
	\title{Quantum Dot Arrays in Silicon and Germanium}

	\author{W. I. L. Lawrie}
	\affiliation{QuTech and Kavli Institute of Nanoscience, Delft University of Technology, Lorentzweg 1, 2628 CJ Delft, The Netherlands}
	\author{H. G. J. Eenink}
	\affiliation{QuTech and Kavli Institute of Nanoscience, Delft University of Technology, Lorentzweg 1, 2628 CJ Delft, The Netherlands}
	\author{N. W. Hendrickx}
	\affiliation{QuTech and Kavli Institute of Nanoscience, Delft University of Technology, Lorentzweg 1, 2628 CJ Delft, The Netherlands}
	\author{J. M. Boter}
	\affiliation{QuTech and Kavli Institute of Nanoscience, Delft University of Technology, Lorentzweg 1, 2628 CJ Delft, The Netherlands}
	\author{L. Petit}
	\affiliation{QuTech and Kavli Institute of Nanoscience, Delft University of Technology, Lorentzweg 1, 2628 CJ Delft, The Netherlands}
	\author{S.~V.~Amitonov}
	\affiliation{QuTech and Kavli Institute of Nanoscience, Delft University of Technology, Lorentzweg 1, 2628 CJ Delft, The Netherlands}
	\author{M.~Lodari}
	\affiliation{QuTech and Kavli Institute of Nanoscience, Delft University of Technology, Lorentzweg 1, 2628 CJ Delft, The Netherlands}
	\author{B. Paquelet Wuetz}
	\affiliation{QuTech and Kavli Institute of Nanoscience, Delft University of Technology, Lorentzweg 1, 2628 CJ Delft, The Netherlands}
	\author{C. Volk}
	\affiliation{QuTech and Kavli Institute of Nanoscience, Delft University of Technology, Lorentzweg 1, 2628 CJ Delft, The Netherlands}
	\author{S. Philips}
	\affiliation{QuTech and Kavli Institute of Nanoscience, Delft University of Technology, Lorentzweg 1, 2628 CJ Delft, The Netherlands}
	\author{G. Droulers}
	\affiliation{QuTech and Kavli Institute of Nanoscience, Delft University of Technology, Lorentzweg 1, 2628 CJ Delft, The Netherlands}
	\author{N.~Kalhor}
	\affiliation{QuTech and Kavli Institute of Nanoscience, Delft University of Technology, Lorentzweg 1, 2628 CJ Delft, The Netherlands}
	\author{F. van~Riggelen}
	\affiliation{QuTech and Kavli Institute of Nanoscience, Delft University of Technology, Lorentzweg 1, 2628 CJ Delft, The Netherlands}
	\author{D.~Brousse}
	\affiliation{QuTech and Netherlands Organization for Applied Scientific Research (TNO), Stieltjesweg 1 2628 CK Delft, The Netherlands}
	\author{A. Sammak}
	\affiliation{QuTech and Netherlands Organization for Applied Scientific Research (TNO), Stieltjesweg 1 2628 CK Delft, The Netherlands}
	\author{L. M. K. Vandersypen}
	\affiliation{QuTech and Kavli Institute of Nanoscience, Delft University of Technology, Lorentzweg 1, 2628 CJ Delft, The Netherlands}
	\author{G. Scappucci}
	\affiliation{QuTech and Kavli Institute of Nanoscience, Delft University of Technology, Lorentzweg 1, 2628 CJ Delft, The Netherlands}
	\author{M. Veldhorst}
	\email{Corresponding Author: m.veldhorst@tudelft.nl}
	\affiliation{QuTech and Kavli Institute of Nanoscience, Delft University of Technology, Lorentzweg 1, 2628 CJ Delft, The Netherlands}
	
	%\date{\today}

	\begin{abstract}
		Electrons and holes confined in quantum dots define an excellent building block for quantum emergence, simulation, and computation. In order for quantum electronics to become practical, large numbers of quantum dots will be required, necessitating the fabrication of scaled structures such as linear and 2D arrays. Group IV semiconductors contain stable isotopes with zero nuclear spin and can thereby serve as excellent host for spins with long quantum coherence. Here we demonstrate group IV quantum dot arrays in silicon metal-oxide-semiconductor (SiMOS), strained silicon (Si/SiGe) and strained germanium (Ge/SiGe). We fabricate using a multi-layer technique to achieve tightly confined quantum dots and compare integration processes. While SiMOS can benefit from a larger temperature budget and Ge/SiGe can make ohmic contact to metals, the overlapping gate structure to define the quantum dots can be based on a nearly identical integration. We realize charge sensing in each platform, for the first time in Ge/SiGe, and demonstrate fully functional linear and two-dimensional arrays where all quantum dots can be depleted to the last charge state. In Si/SiGe, we tune a quintuple quantum dot using the N+1 method to simultaneously reach the few electron regime for each quantum dot. We compare capacitive cross talk and find it to be the smallest in SiMOS, relevant for the tuning of quantum dot arrays. These results constitute an excellent base for quantum computation with quantum dots and provide opportunities for each platform to be integrated with standard semiconductor manufacturing. 
	\end{abstract}
	\maketitle
	
	Quantum dots have been a leading candidate for quantum computation for more than two decades \cite{Loss1998}. Furthermore, they have matured recently as an excellent playground for quantum simulation \cite{Hensgens2017} and have been proposed for the design of new states of matter \cite{Leijnse2011, Sau2012}. Pioneering studies in group III-V semiconductors led to proof-of-principles including the coherent control of electron spins \cite{Petta2005,Koppens2006}, rudimentary quantum simulations \cite{Dehollain2019}, and signatures of Majorana states \cite{Mourik2012}. The group IV semiconductors silicon and germanium have the opportunity to advance these concepts to a practical level due to their compatibility with standard semiconductor manufacturing \cite{Vandersypen2017} and the availability of isotopes with zero nuclear spin, increasing quantum coherence for single spins by four orders of magnitude \cite{Veldhorst2014}. Furthermore, heterostructures built from silicon and germanium may offer a large parameter space in which to engineer novel quantum electronic devices \cite{Zwanenburg2013b,Sammak2019,Sabbagh2018}.
	
	An initial advancement towards silicon quantum electronics \cite{Zwanenburg2013b} was the design of an integration scheme based on overlapping gates to build silicon metal-oxide-semiconductor (SiMOS) quantum dots \cite{Angus2007a}. This technique was later adopted in strained silicon (Si/SiGe) \cite{Zajac2015} and refined by incorporating metals with small grain size and atomic layer deposition (ALD) for layer-to-layer isolation \cite{Brauns2018} and to enable tunable coupling between single electrons in SiMOS \cite{eenink2019}. These developments in fabrication have led to a great body of results, including high-fidelity qubit operation \cite{Yang2019a,Yoneda2018} and two-qubit logic \cite{Veldhorst2015,Zajac2018,Watson2018}. Controlling holes in silicon has been more challenging due to type II band alignment in strained silicon, limiting experiments to SiMOS \cite{Spruijtenburg2013, Liles2018, Maurand2016}. Strained germanium on the other hand \cite{Failla2016,Su2017,Sammak2019} exhibits type I band alignment and is thereby a viable platform in which holes with light effective mass \cite{Lodari2019} can be confined \cite{Hendrickx2018} and coherently controlled \cite{Hendrickx2019a}. This motivates the development of an integration scheme that can build upon the individual breakthroughs realized in each platform to advance group IV semiconductor quantum dots towards large quantum systems.
	
	Here, we present the fabrication and operation of quantum dots in silicon and germanium, in linear and two-dimensional arrays. We show stability diagrams obtained by charge sensing and report double quantum dots in SiMOS, Si/SiGe, and Ge/SiGe that can be depleted to the last charge state. We compare integration schemes and find that while each platform has unique aspects and opportunities, the core fabrication of overlapping gates defining the nano-electronic devices is remarkably similar. Fabrication is most demanding in SiMOS due to requirements on feature size, but we also find that the resulting devices have the smallest cross capacitance, simplifying tuning and operation. We leverage off the ohmic contact between quantum dots in Ge/SiGe and metals \cite{Dimoulas2006} to avoid the need for implants and to provide means for novel hybrid systems. In each case, fabrication starts from a silicon substrate, and integration is compatible with standard semiconductor technology.
	
	Figure 1a schematically shows the SiMOS, Si/SiGe, and Ge/SiGe wafer stacks used in this study. The SiMOS 300 mm wafers are grown in an industrial complementary metal-oxide-semiconductor (CMOS) fab \cite{Petit2018,eenink2019,Sabbagh2018}, while the Si/SiGe and Ge/SiGe four-inch wafers are grown using an RP‐CVD reactor (ASM Epsilon 2000) \cite{Sammak2019}. Each platform is grown on a p-type natural Si wafer. The SiMOS structure consists of 1 $\mu$m intrinsic natural silicon (${}^i$Si)  followed by 100 nm ${}^{28}$Si (800 ppm purity) and 10 nm SiO$_2$ \cite{Sabbagh2018}. The Si/SiGe heterostructure begins with a linearly graded Si${}_{1-x}$Ge${}_{x}$ layer, where x ranges from 0 to 0.3. A relaxed Si${}_{0.7}$Ge${}_{0.3}$ layer of 300 nm lies below the 10 nm ${}^{28}$Si (800 ppm purity) quantum well which itself is separated from the 2 nm Si capping layer by a second 30 nm relaxed Si${}_{0.7}$Ge${}_{0.3}$  spacer layer. The Ge/SiGe wafer stack starts with 1.4 $\mu$m of Ge and 900 nm of reverse graded Si${}_{1-x}$Ge${}_{x}$ where x ranges from 1 to 0.8. This lies below a 160 nm  Si${}_{0.2}$Ge${}_{0.8}$ spacer layer, a 16 nm Ge quantum well under compressive strain, a second Si${}_{0.2}$Ge${}_{0.8}$ layer of 22 nm and finally a thin Si cap of 1 nm \cite{Sammak2019}.\\%
	
	Figure 1b shows a carrier mobility versus density characterization of the three platforms. Hall bar structures were fabricated on coupons cut from the center of each wafer. Maximum mobility and critical density are extracted at 1.7~K. SiMOS 300 mm processed wafers give a peak mobility value of 1$\times 10^{4}$ cm${}^2/$Vs, as well as a critical density of about 1.75$\times 10^{11}$ cm${}^{-2}$ as shown in another work \cite{Sabbagh2018}. At higher densities, SiMOS mobilities fall off due to surface roughness scattering effects \cite{Ando1982,Gold1986,Kruithof1991}. In Si/SiGe, we observe a lower critical density of 1.2$\times 10^{11}$ cm${}^{-2}$ and a significantly higher maximum mobility exceeding 1$\times 10^5$ cm${}^2/$Vs . Similar studies conducted on natural Si/SiGe grown in an industrial CMOS fab yielded mobilities of 4.2$\times 10^{5}$ cm${}^2/$Vs \cite{Wuetz2019}. This quality improvement observed by moving toward industrial CMOS fab also suggests encouraging prospects for Ge/SiGe, already exhibiting a high maximum mobility of 5$\times 10^5$ cm${}^2/$Vs and critical density of 1.15$\times 10^{11}$ cm${}^{-2}$ despite being grown in an academic cleanroom via RP-CVD \cite{Sammak2019}.\\
	
	\begin{figure}
		\centering
		\includegraphics[]{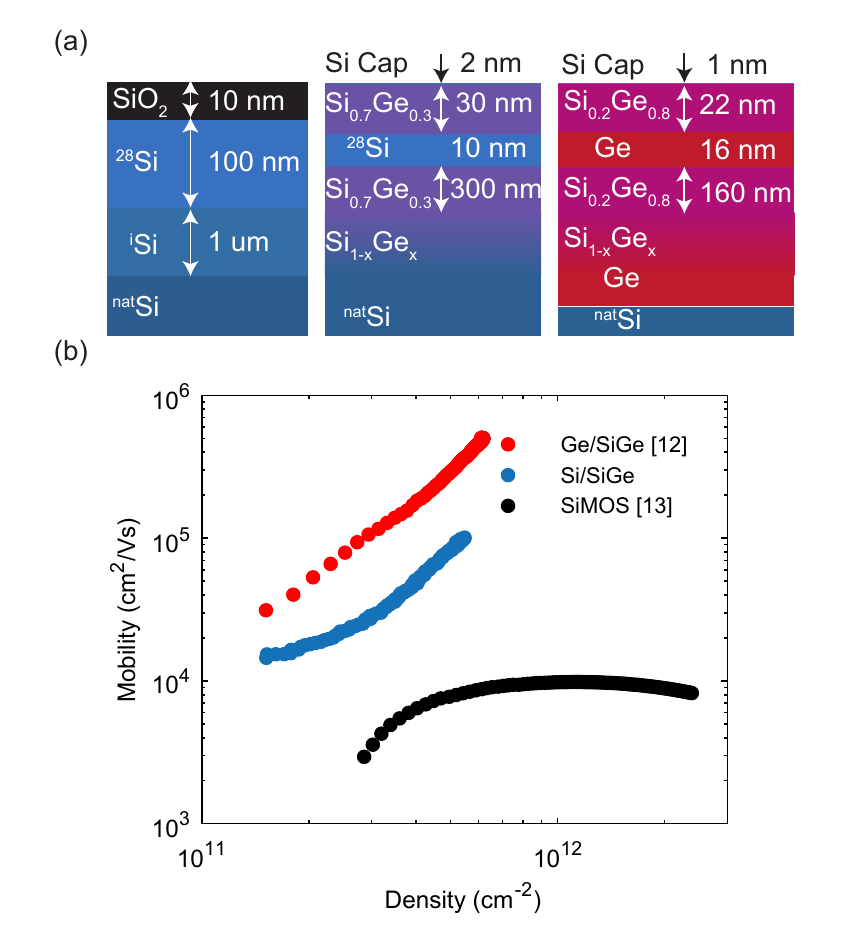}
		\caption{Wafer stack schematics and mobility as a function of carrier density. (a) From left to right, SiMOS, Si/SiGe, and Ge/SiGe wafers stacks. For SiMOS, a ${}^{28}$Si epilayer with 10 nm thermal oxide is grown on a 1 $\mu$m intrinsic natural Si buffer layer. The Si/SiGe heterostructure consists of a 1.5 $\mu$m linearly graded SiGe layer, a relaxed 300 nm SiGe spacer, a 10 nm ${}^{28}$Si quantum well, a 30 nm SiGe spacer, and a 2 nm Si cap. The Ge/SiGe heterostructure consists of  900 nm  reverse graded SiGe layer, a relaxed 160 nm SiGe spacer, a 16 nm Ge quantum well, a 22 nm SiGe spacer, and a 1 nm Si cap. (b) Mobility as a function of carrier density measured in each platform. For Ge/SiGe, the peak mobility is greater than 5$\times 10^5$ cm${}^2$/Vs and the critical density is  1.15$\times 10^{11}$ cm${}^{-2}$ \cite{Sammak2019}. The same measurements for Si/SiGe wafers give a peak mobility of 1$\times 10^5$ cm${}^2$/Vs and a critical density of 1.2$\times 10^{11}$ cm${}^{-2}$. SiMOS data taken from \cite{Sabbagh2018} shows a mobility of 1$\times 10^4$ cm${}^2$/Vs and a higher critical density of 2.5$\times 10^{11}$ cm${}^{-2}$. }
		
		\label{fig:fig1}
	\end{figure}
	\begin{figure}[t!]
		\centering
		\includegraphics[width=\linewidth]{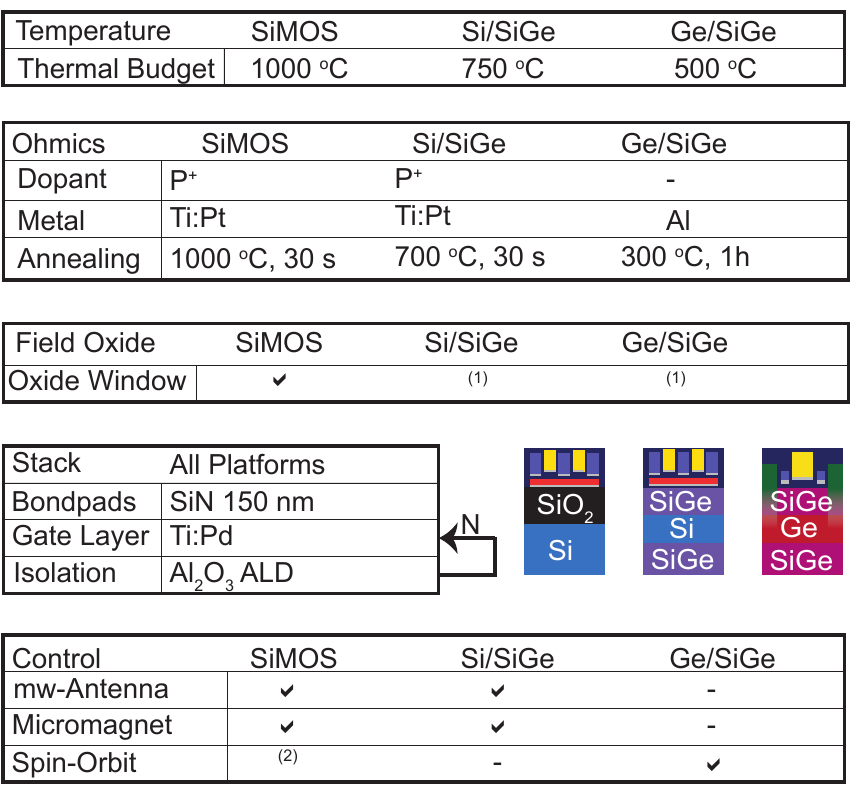}
		\caption{Overview of fabrication scheme for SiMOS, Si/SiGe and Ge/SiGe quantum dots. The thermal budget of each material prior to gate stack deposition is estimated based on the limiting mechanism of each platform as discussed in the text. In all cases, gates are fabricated from Pd metal with a thin (3 nm) Ti adhesion layer, with layer-to-layer isolation performed via atomic layer deposition (ALD) of Al${}_2$O${}_3$. These two steps can be looped at appropriate thicknesses to form the multi-layer structure. (1) We note the possibility of such an etch exists for the remaining platforms in the case of a Schottky gate architecture (2) We note that spin-orbit based driving of electrons in SiMOS has been demonstrated for singlet-triplet qubits \cite{Jock2018} and proposed for single spin qubits \cite{Huang2017}. }
		\label{fig:fig2}
	\end{figure}
	
	\begin{figure*}[t!]
		\centering
		\includegraphics[width=\textwidth]{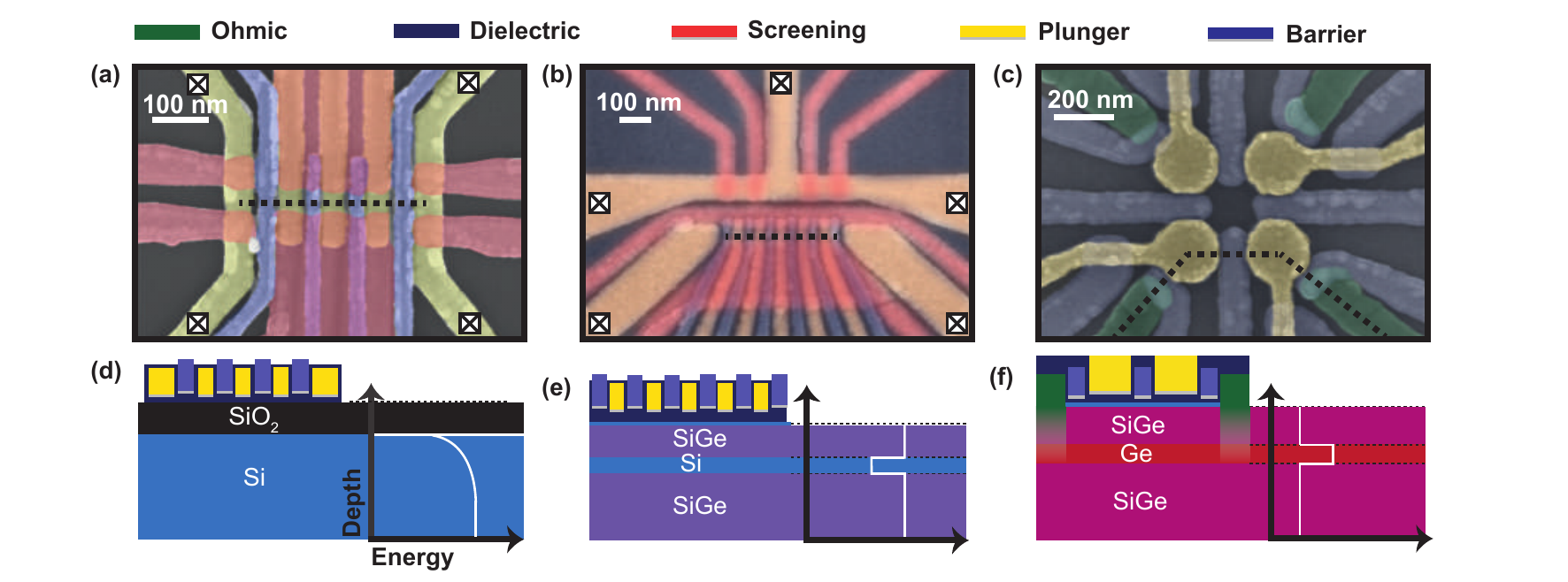}
		\caption{Scanning electron microscope images and corresponding substrate with band bending diagrams and gate stack for each of the devices. Dotted lines in (a-c) indicate the cross-section through the dot channel illustrated in (d-f) respectively, and crossed boxes indicate gates that overlap with implanted regions to form ohmic contact. The plunger gates (yellow), the barrier gates (blue) and the screening gates (red) define the quantum dots. (a) SiMOS triple quantum dot linear array. Two SETs function as charge sensors and as reservoirs for the quantum dots on either side of the array  (b) Si/SiGe quintuple quantum dot linear array. Two SETs (top) are used for charge sensing. (c) Ge/SiGe (2x2) quadruple quantum dot array. Each dot is tunnel coupled to a metallic lead (green). Measurement can be performed in transport, or using charge-sensing by forming a sensor by coupling two quantum dots. (d,e,f) Cross-section and bandstructure of metal, dielectric (black) and semiconductor (d) SiMOS, (e) Si/SiGe and (f) Ge/SiGe.
		}
		\label{fig:fig3}
	\end{figure*}
	
	Figure 2 summarises the integration scheme utilized for each platform. The thermal budget is estimated based on the respective limiting mechanisms. For SiMOS, thermal processing is limited by the self-diffusion of natural silicon from the substrate into the ${}^{28}$Si epilayer. From the self diffusion constants measured by Bracht et al. \cite{Bracht1998}, we estimate the point at which the residual ${}^{29}$Si concentration within 1 nm of the Si-SiO${}_2$ interface increases by 1 ppm occurs at 1000 ${}^o$C for time scales above 1 hour, for furnace anneals in a pure argon atmosphere.
	Consequently, this allows for extensive thermal treatment and annealing of samples. This is highly advantageous, as we have observed that a 15 minute anneal in forming gas at 400 ${}^o$C after the deposition of every gate layer greatly improves the quality of our fine features (see Supporting Information section Ia for detailed comparison). 
	In addition, a final end-of-line anneal is conducted to eliminate processing damage at 400 ${}^o$C in forming gas for 30 minutes. In the cases of Si/SiGe and Ge/SiGe, the thermal budget is limited by strain relaxation of the quantum wells, thus the maximum processing thermal budget is given qualitatively by the temperature at which the quantum wells were grown. This is 750 ${}^o$C for strained Si and 500 ${}^o$C for strained Ge \cite{Sammak2019}.
	
	The design of ohmic contacts is tailored to the specific requirements of the device. For both Si platforms, ohmic contact is made via high fluence P ion implantation followed by evaporation of Ti:Pt metallic contacts, creating n$^{++}$ doped, low resistance channels. The oxide is etched locally directly before metal deposition using buffered hydrofluoric acid (BHF). In the case of Si/SiGe, stray capacitance is minimized to ensure maximum power is dissipated in the variable resistance of the sensing quantum dot for RF-readout. Germanium can make direct ohmic contact to metals \cite{Dimoulas2006}, avoiding the need for implants. We deposit Al and anneal at 300 ${}^o$C for 1 hour in vacuum to assist in Al diffusion into the quantum well. The Al ohmic is defined close to the quantum dots, resulting in a very low resistance channel ideally suited for RF circuits and enabling a tunnel contact that can even be made superconducting \cite{Hendrickx2019}. The implementation does however lower the thermal budget of further processing.\\
	
	Fabrication of each device utilizes a titanium-palladium (Ti:Pd) gate stack with 3 nm of Ti deposited for each layer to assist with adhesion. Pd makes a good gate metal due to its low grain size \cite{Brauns2018}. Unlike the commonly used material Al, Pd does not self-oxidise and ALD can be used to define sharp dielectric interfaces. For the SiMOS and Si/SiGe devices shown in Fig. 3, we utilize a three layer gate stack that we refer to as the screening layer, the plunger layer and the barrier layer. In order to assist climbing of overlapping gate features, the initial layer is deposited at 20 nm total thickness, while subsequent layers at 40 nm. Each layer is isolated from one another via ALD of Al${}_2$O${}_3$ at 7 nm thickness. We measure the dielectric strength of our Al${}_2$O${}_3$ to be  greater than 6 MV/cm, allowing potentials of greater than 4 V to be applied between adjacent gates. To leverage off the high quality industrial CMOS fab, we begin fabrication of SiMOS devices on wafers including a 10 nm SiO${}_2$ oxide already grown. To further reduce likelihood of leakage from gate to substrate, we first grow a thick 10 nm Al${}_2$O${}_3$ blanket layer over the entirety of the substrate. Advantageously, one can etch Al$_2$O$_3$ on thermally grown SiO$_2$ selectively, allowing the definition of a 20x20 $\mu$m${}^2$ area where the quantum dot system is defined, which we have measured to significantly reduce low-frequency drifts deduced from charge occupation stability \cite{Connors2019} (see Supporting Information section Ib for comparison). \\
	
	\begin{figure*}[t!]
		\centering
		\includegraphics[]{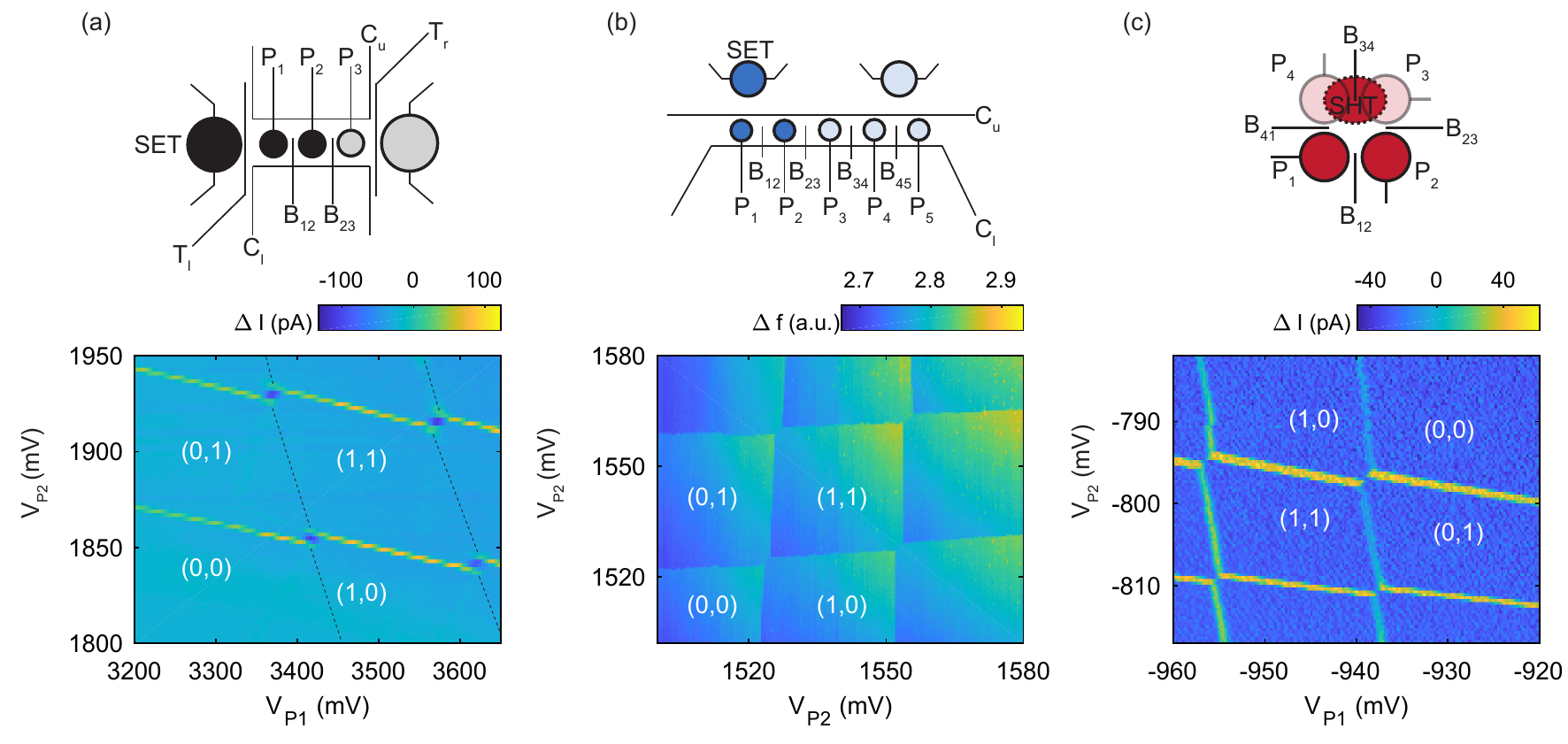}
		\caption{Charge stability diagrams of double quantum dots depleted to the single electron/hole regime for the three platforms. (a) SiMOS double quantum dot. Charge addition lines under P${}_1$ are not visible due to low tunnel rate from reservoir. Map taken at 0.44 K using lock-in  charge sensing. The excitation is placed on the inter-dot gate B${}_{12}$.  (b) Si/SiGe double quantum dot formed under the first two plungers, sensed by the nearest charge sensor via RF-reflectometry utilizing a resonant LC circuit at 84 MHz. Here, the plunger gate voltages are in virtual gate space correcting for weak cross capacitive coupling. (c) Ge/SiGe depleted to the single hole regime. A large single dot is formed under P${}_3$, B${}_{34}$ and P${}_4$, by adjusting the tunnel barrier voltage B${}_{34}$, and is used to sense a double quantum dot under P${}_1$ and P${}_2$. The lock-in excitation is placed on the inter-dot tunnel barrier B${}_{12}$. }
		\label{fig:fig4}
	\end{figure*}
	\begin{figure*}[t!]
		\centering
		\includegraphics[]{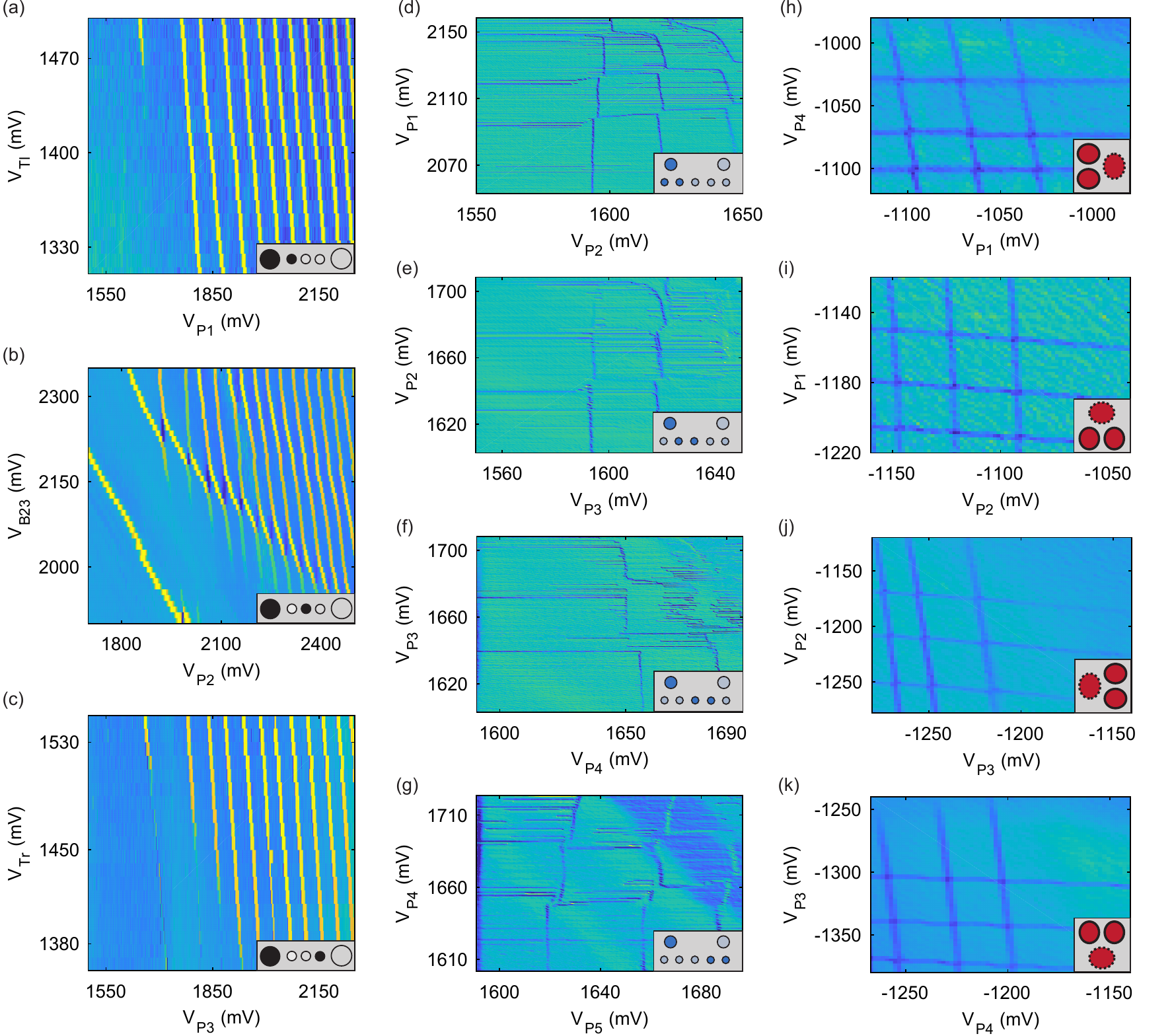}
		\caption{Quantum dot arrays in SiMOS, Si/SiGe and Ge/SiGe. (a-c) SiMOS triple dot device stability diagrams. Each single dot is formed under its respective plunger gate upon which an excitation is placed for lock-in charge sensing. Each dot is depleted to the single charge state. (b) Shows the crossing of the adjacent dot under P${}_3$, through which the dot is loaded. (d-g) Si/SiGe double dots tuned up sequentially using the N+1 method \cite{Volk2019} to the single electron regime. True plunger gate voltages are plotted, though virtual gates are swept containing small corrections to adjacent barriers and plungers. Each double dot pair is sensed using RF-reflectometry. The same SET is used for readout in each case, as indicated by the relative signals as each double dot pair is formed farther from the charge sensor. (g) The data has been filtered to remove 50 Hz background noise for data clarity.  (h-k) Ge/SiGe 2x2 array double dots formed in each possible configuration. In each case, a charge sensor is formed in the parallel channel by raising the inter-dot coupling to form a large single dot with high hole occupation. Each charge stability diagrams shows RF-sensing of double quantum dots depleted to the last hole occupancy, in the low tunnel-coupled regime. }
		\label{fig:fig5}
	\end{figure*}
	The final deposition step is the qubit control layer. The spin-orbit coupling for holes in germanium enables qubit operation by simply applying microwave pulses to the quantum dot gates \cite{Watzinger2018, Hendrickx2019a} and no further processing is required. In silicon, qubit driving can be realized by integrating on-chip striplines \cite{Veldhorst2014}, which we fabricate using Al or NbTiN, or micromagnets \cite{Kawakami2014}, which we integrate using Ti:Co. Quantum dots in Si/SiGe generally have a larger and more mobile electron wave function as compared to SiMOS and thereby benefit most from a micromagnet integration for fast qubit driving.\\
	
	A schematic of each material and associated device is shown in Fig. 3 and labelling of the relevant gates are shown in Fig. 4. The SiMOS device is a three-layer, triple quantum dot structure with dedicated plungers (P${}_{1-3}$), inter-dot barriers (B${}_{12}$, B${}_{23}$)  and reservoir barriers (T${}_l$, T${}_r$). Charge noise resulting from fluctuations of impurities near the quantum dot array is screened by two large metallic gates (C${}_\text{l}$, C${}_\text{u}$) deposited in the initial layer and kept at constant potential. These also serve to confine the quantum dots in one lateral dimension. Two single electron transistors (SETs) are positioned at either side of the quantum dot array, and function as charge sensors for spin and charge readout.
	The Si/SiGe device is a quintuple quantum dot linear array written in three layers utilizing a similar architecture to that of the SiMOS device. The dot array contains five plunger gates (P${}_{1-5}$) with inter-dot barriers (B${}_{12-45}$) and reservoir barriers. Dots are confined laterally and screened from charge noise by two confinement gates. Two SETs are positioned parallel to the dot channel. The Ge/SiGe device is a 2x2 quadruple quantum dot array written in two layers. Gates (P${}_{1-4}$) are positioned anti-clockwise in the array and define the potential of the dots. Each pair of adjacent dots share a barrier gate (B${}_{12-41}$) capable of tuning inter-dot tunnel coupling. Coupling of each dot to its reservoir can be controlled via a barrier gate. This device can be operated as a quadruple quantum dot system in transport mode, but for the present work we intentionally tune the inter-dot barrier to form a single hole transistor (SHT) along a dot channel that we subsequently use for charge sensing of the double quantum dot along the opposite channel. 
	For more information about device specific fabrication, see Supporting Information section II.
	
		\begin{figure}[h!]
		\centering
		\includegraphics[]{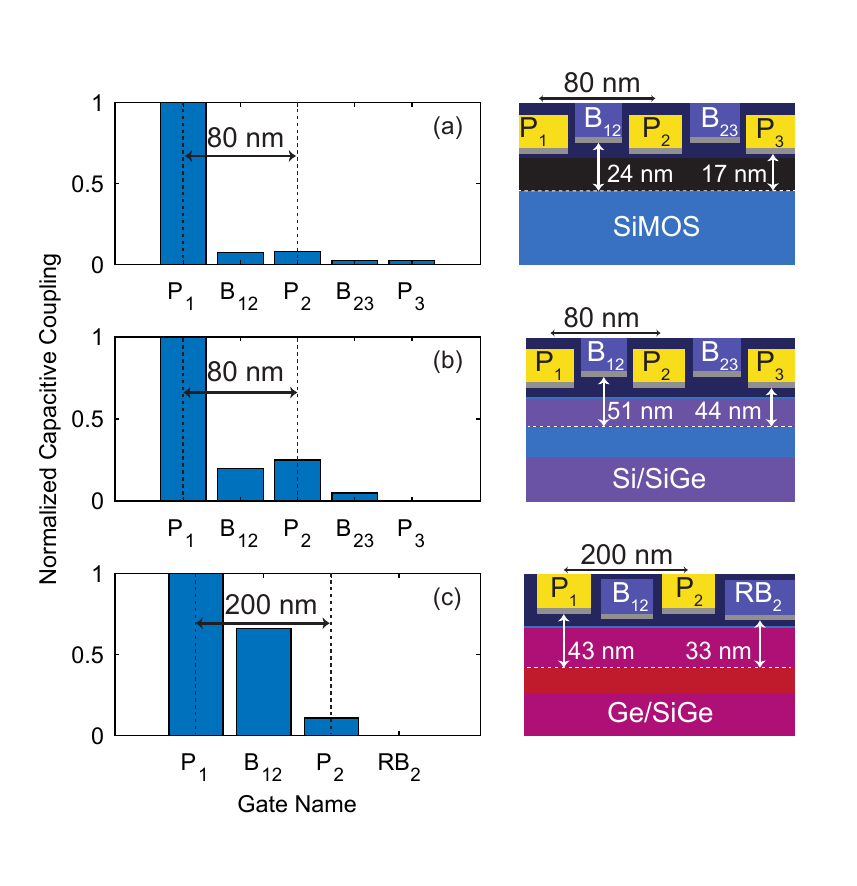}
		\caption{Cross capacitance to neighbouring gates of a quantum dot in the single charge occupancy regime under gate P${}_1$ in each platform. For SiMOS (a), we observe immediate falloff of cross coupling due to the tight dot confinement present in SiMOS devices. Here the inter-dot pitches matches that of Si/SiGe at 80 nm. For Si/SiGe (b), we see significant cross-coupling between adjacent plungers and barrier gates. Here the plunger gates are written before the barrier layer and have an inter-dot pitch of 80 nm. Ge/SiGe (c) reveals as expected a slower fall-off of cross coupling. We attribute this to the larger plunger gate design, made possible by lower hole effective mass. In this case, the plunger gates P${}_1$ and P${}_2$ are written in the layer above the barrier gates B${}_{12}$ and RB${}_2$, decreasing coupling to their respective quantum dots. The plunger to plunger pitch is 200 nm. Each cross-sectional cartoon shows plunger pitch and distance between each relevant gate layer to the center of the quantum well. }
		\label{fig:fig6}
	\end{figure}
	To demonstrate the success of our integration scheme, we show that we can create stable quantum dots in each platform. Figure 4 shows charge stability diagrams for tunnel-coupled double quantum dots, measured by performing charge sensing. Lock-in techniques are used in the case of SiMOS and Ge/SiGe, where an excitation is placed on an inter-dot barrier gate B${}_{12}$ in each case, and the trans-conductance of our source-drain channel is measured. We use compensation to remain at a sensitive point on our SET/SHT Coulomb peaks \cite{Yang2011}. In the case of Si/SiGe, charge readout is performed using RF-reflectometry techniques. A 3 $\mu$H kinetic inductor is bonded to the sample source which forms a resonant LC circuit when combined with parasitic capacitance to ground. In each case we measure a stability diagram and show that we can deplete down to the (0,0) electron/hole charge configuration. We note that the plunger voltages in the case of Si/SiGe required to form double quantum dots in the (1,1) charge occupation are within a charging energy. These remarkably similar tuning parameters are promising with regards to the stringent requirements placed on quantum dot array tune-up in crossbar architectures \cite{Li2018}. While operation in the single electron regime in silicon has been routinely achieved before, this work shows the first demonstration of the single hole regime using charge sensing of holes in Ge/SiGe. We attribute the slight difference in slope of the first and second charge addition lines in Fig. 4c to a shift in the position of the dot relative to the inter-dot tunnel barrier.\\
	
	In Fig 5. we demonstrate that quantum dots can be formed under each dedicated plunger gate. For Fig. 5 (a-c), in each SiMOS quantum dot, lock-in charge sensing is performed by placing an excitation on the respective plunger gates, while trans-conductance in the nearby SET channel is measured. In each case, the first charge transition is visible. For quantum dots formed under plungers P${}_{2-3}$, electron loading is from the right SET which constitutes a reservoir. For the dot under P${}_1$, loading is from the left SET via the gate T${}_l$. The Si/SiGe quintuple quantum dot system in Fig 5(d-g) is tuned using the N+1 strategy \cite{Volk2019}, reaching the few-electron regime  simultaneously for all quantum dots. In Fig. 5 we show stability diagrams, in each of which we scan two virtual plunger gates which allow to controllably load a single electron into each quantum dot. Double quantum dots are formed between each set of adjacent plungers, and sensed using RF-reflectometry like in Fig. 4b using the left SET for all configurations. As expected, observable signal from charge transition lines fades as the dot pairs are formed farther away from the SET. The derivative of the reflected signal is plotted, and shows the (0,0) charge occupancy for each charge stability diagram. For every double dot, loading occurs via the left accumulation gate, leading to latching effects and low tunnel rates in the dots formed farther away from the reservoir. Here, the plunger voltages, while similar, are not entirely within a charging energy, suggesting further improvements to heterostructure uniformity are required to meet strict large scale array tune-up requirements. Figure 5 (h-k) shows charge sensing operation of the 2x2 quantum dot array fabricated in Ge/SiGe. In each case, a sensing dot is formed in the channel parallel to the double dot by opening the inter-dot barrier such that a large single dot is formed. In the opposite channel, the inter-dot barrier is closed, forming a double dot system in the low tunnel coupled regime.    \\
	
	A significant challenge for larger quantum dot arrays will manifest in tuning. The presence of large capacitive crosstalk in GaAs has led to development of virtual gates and approaches to tune larger systems \cite{Baart2016,Volk2019}. To assess the relevance of these approaches for silicon and germanium structures we measure cross capacitance as show in Fig. 6. To obtain the cross coupling, we measure the slope of the charge addition lines with respect to each gate and normalize by a cross coupling of unity for the plunger gate associated with the respective quantum dot. Each slope is taken for the first charge transition and in the low tunnel-coupled regime. In SiMOS, cross coupling is almost negligible, as expected from quantum dots located only 17 nm (10 nm SiO$_2$ and 7 nm Al${}_2$O$_3$) below the electric gates. This compares favourably to the cross coupling observed in Si/SiGe, where falloff is significantly slower despite sharing equal gate pitch to the SiMOS array. While the cross coupling in the Ge/SiGe system is the largest and extends over multiple neighbouring gates, it still falls off significantly faster than quantum dots defined in GaAs \cite{Volk2019}. For Ge/SiGe, we also observe that the barrier gates have a relatively stronger coupling as compared to the plunger gates, due to definition in lower layers of the multi-layer stack. Summarizing, we conclude that for SiMOS tuning is most straightforward considering capacitive cross talk only, while each platform compares favourably to GaAs.\\
	
	In conclusion we presented a cross-platform integration scheme for multi-layer quantum dot arrays in group-IV semiconductor hosts. We successfully fabricated linear and 2D arrays of quantum dots and in the group IV platforms SiMOS, Si/SiGe and Ge/SiGe. We demonstrated single electron and hole occupancy in double quantum dots confirmed by charge sensing. We showed stable quantum dots under each plunger in a SiMOS triple dot linear array, depleteable to the final charge state. In Si/SiGe, we demonstrated tune-up of a quintuple dot array utilizing the N+1 method, successfully reaching the few electron regime in each dot simultaneously. Moreover, we showed we could form and sense double dots in the single hole regime in each configuration of a 2x2 quadruple quantum dot array in Ge/SiGe. We furthermore compared the capacitive cross talk between quantum dots and gates. We find that the cross capacitance can be small and therefore argue that future work on strategies for the initial tuning of quantum dot arrays should address disorder rather than capacitive cross talk, in particular for SiMOS quantum dots. We envision that our realization of an integration scheme to build quantum dots in SiMOS, Si/SiGe,
	and Ge/SiGe will boost the collective development toward large quantum dot arrays to build, simulate, and compute with quantum information.

	. 
	\section*{Acknowledgements}
	M.V. acknowledges funding by the Netherlands Organization of Scientific Research (NWO) for a VIDI grant and for a projectruimte. The authors are grateful for support from Intel. Research was sponsored by the Army Research Office (ARO) and was accomplished under Grant No. W911NF- 17-1-0274. The views and conclusions contained in this document are those of the authors and should not be interpreted as representing the official policies, either expressed or implied, of the Army Research Office (ARO), or the U.S. Government. The U.S. Government is authorized to reproduce and distribute reprints for Government purposes notwithstanding any copyright notation herein. 
	
\providecommand{\refin}[1]{\\ \textbf{Referenced in:} #1}

\clearpage
\newpage
\onecolumngrid

\renewcommand{\figurename}{Supporting Figure}
\setcounter{figure}{0}  
%\appendix*
\begin{center}
	\textbf{\large Supporting Information}
\end{center}
\renewcommand{\bibnumfmt}[1]{[S#1]}
\renewcommand{\citenumfont}[1]{S#1}
	\section{SiMOS Fabrication Improvements}
\subsection{Gate Anneal}
In silicon MOS, anneals are generally used to repair damage caused by e-beam exposure and to improve the structural quality of metal gates.
We find that for SiMOS quantum dot devices, the quality of the gates can be improved by the incorporation of a forming gas anneal at 400 $^{\circ}$C for 15 min after each gate deposition. Supporting Figure 1(a) shows a scanning electron microscope (SEM) image of a device for which the anneal was implemented. Supporting Figures 1(b)-(g) show SEM and Atomic Force Microscopy (AFM) images for two gate layers of a SiMOS device. We observe a large reduction of surface roughness and sidewall height, which improves even further the homogeneity and yield of the metallic gates.

\begin{figure}%
	\includegraphics[width=\linewidth]{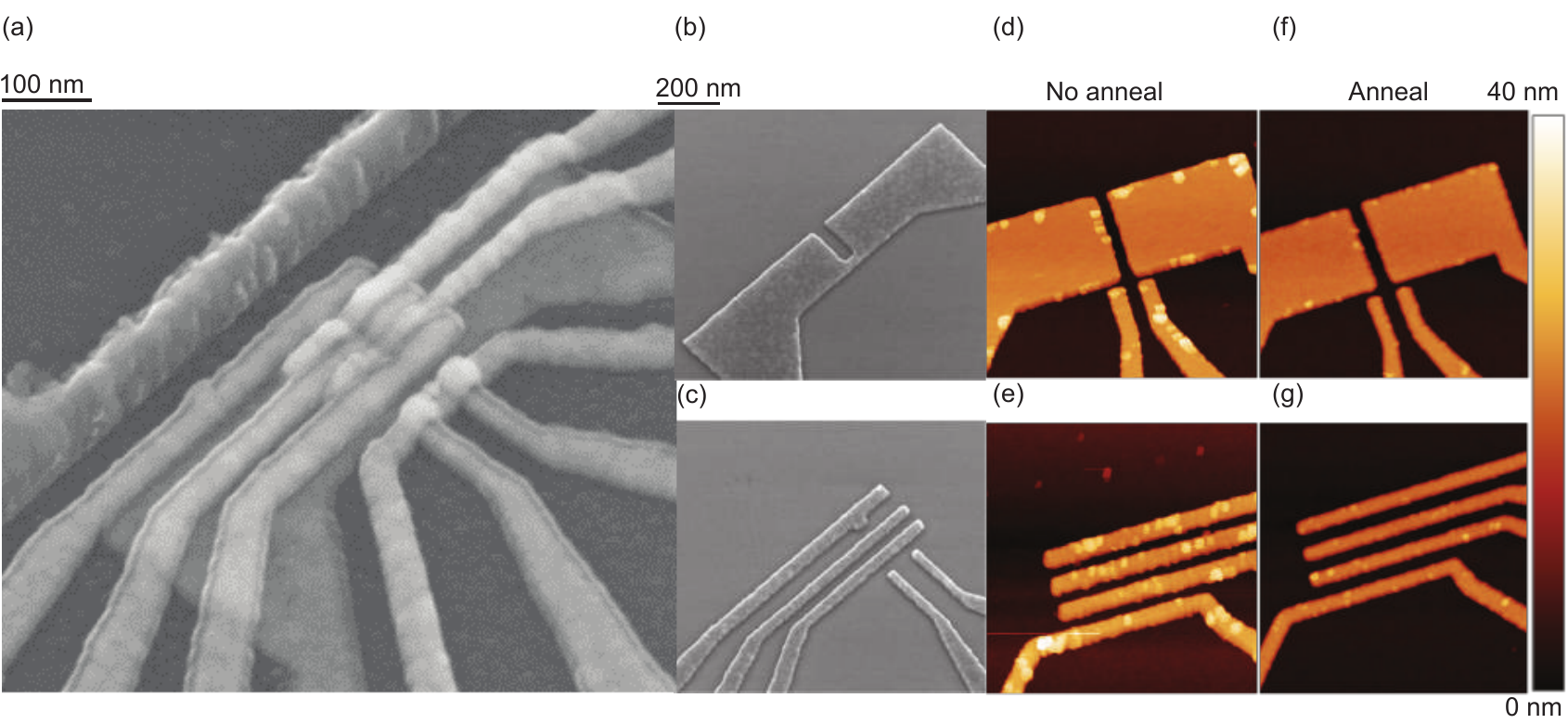}%
	\caption{(a,b,c) SEM images taken under a 30$^{\circ}$ angle of (a) full device fabricated with a gate anneal. (b,c) Separate layers after a gate anneal. (d,e,f,g) AFM images of separate gate layers before (d,e) and after (f,g) a gate anneal. The anneal results in a smoother surface with less grains and sidewalls.}
	\label{fig:ald}
\end{figure}
\subsection{ALD Window Etch}
Incorporating atomic layer deposition (ALD) of Al$_2$O$_3$ into the gate stack introduces further sources of charge noise \cite{Connors2019} making it undesirable in the active region of spin qubits. On the other hand, it is necessary to prevent inter-gate leakage when utilizing a Ti:Pd multi-layer gate stack, as well as leakage to substrate. In the case of SiMOS, fabrication begins on a 10 nm SiO${}_2$ dielectric grown across the substrate. While high in quality, due to the sheer area overlap of gate fan-out, there is a non-negligible probability that a gate may overlap with a region of damaged dielectric.
To prevent leakage of gate layers to substrate in our SiMOS stack, we find an initial blanket layer of Al$_2$O$_3$ is necessary.
An etching process (see Supp. Info. section II regarding fabrication details) with a high selectivity of Al$_2$O$_3$ over SiO$_2$ allows us to locally remove this layer in the active region.
Supporting Figure 2 shows charge stability diagrams of two identically processed quantum dot devices in SiMOS where Al$_2$O$_3$ was present or where an oxide window was etched. Without etching, charge noise causes significant fluctuations in the quantum dot potential, which can be observed from the constantly shifting charge addition lines in Supporting Fig. 2a. Instead, when an oxide window is etched, we observe stable transitions, see Supporting Fig. 2b. We attribute this stability to the removal of the ALD layer beneath the first gate layer.  We note that this behaviour is reproducible in and consistent with other SiMOS quantum dot devices fabricated with and without the removal of the initial ALD layer in the quantum dot active region. 
\begin{figure}%
	\includegraphics[width=\linewidth]{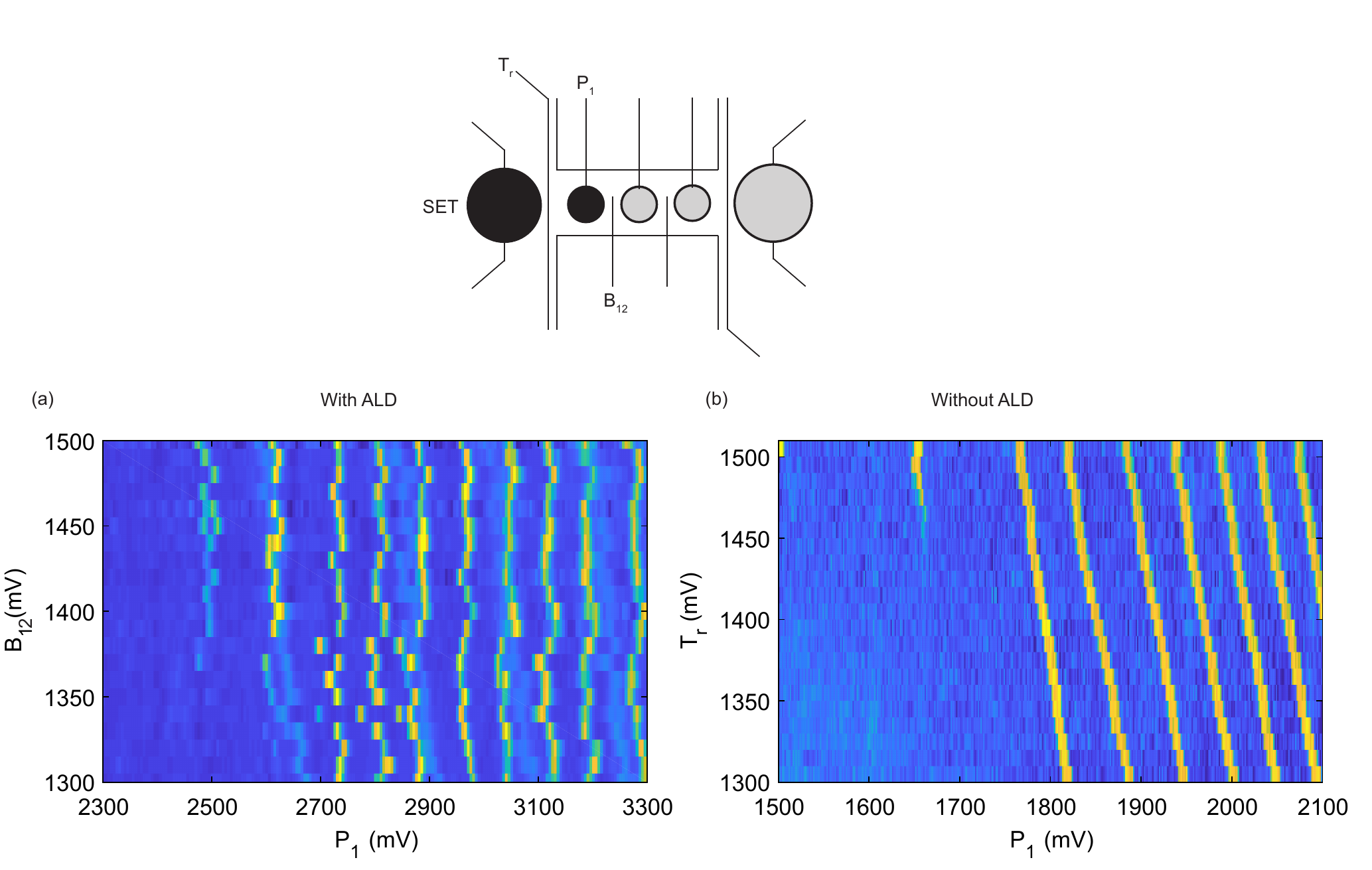}%
	\caption{Charge stability diagrams via lock-in charge sensing of identically processed quantum dot devices in SiMOS, with the exception of an oxide window etch step. (a) Typical stability behaviour of a single quantum dot with a layer of Al${}_2$O${}_3$ beneath the screening layer. (b) Device processed with oxide window etch step. Stability markedly improves over previous case.  }
	\label{fig:ald}
\end{figure}
\newline
\newpage
\section{Extended Fabrication Recipe}

Here we present the full fabrication process for each platform, specifically for the devices fabricated and studied in the main text. \\
\newline
\begin{figure*}[h!]%
	\includegraphics[width=\linewidth]{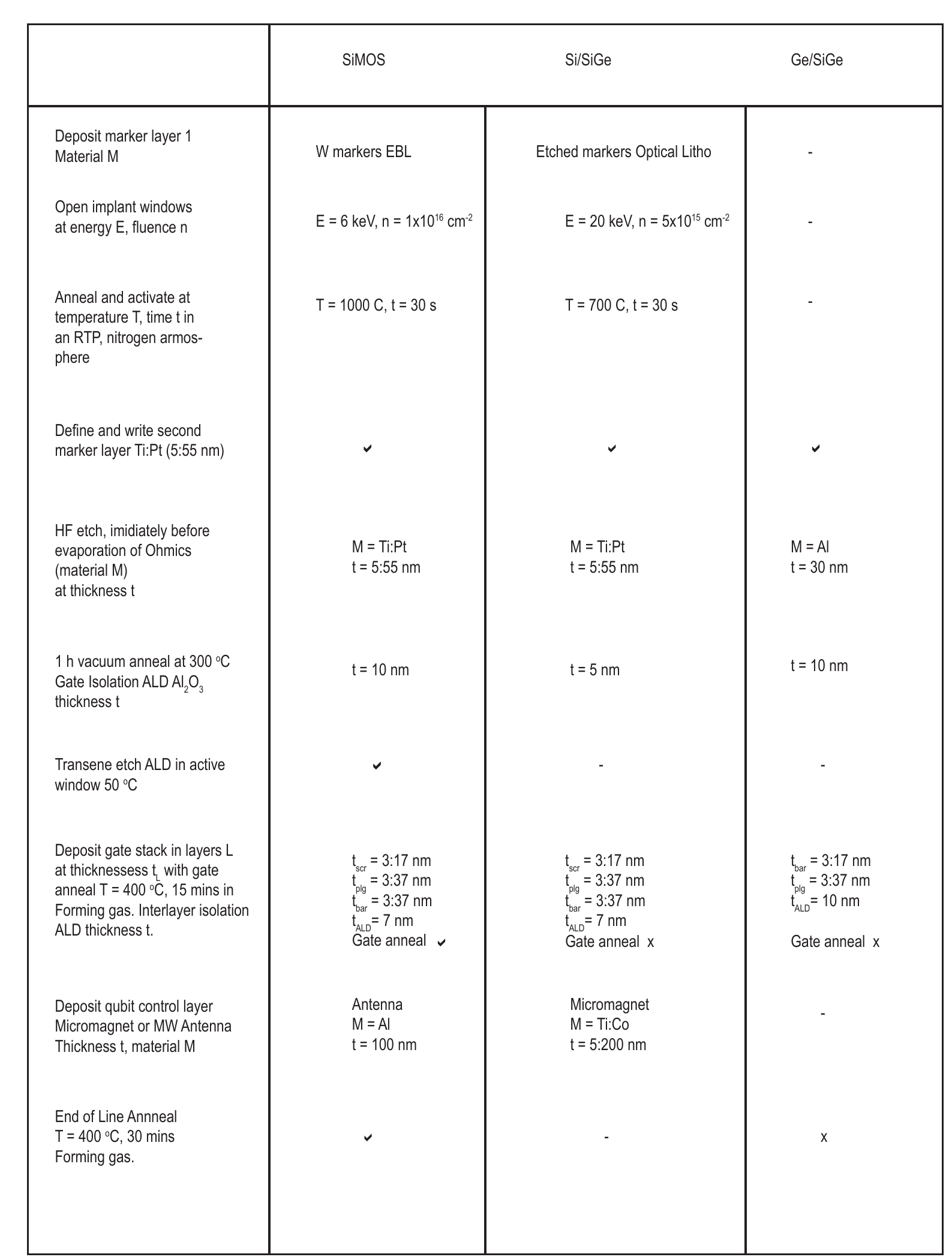}%
	\caption{Full fabrication recipe for the three platforms presented in this work. Ticks represent a used and compatible process, dashes represent a not-used but not nessecarily incompatible process, and crosses indicate an unused and incompatible process. }
	\label{fig:ald}
\end{figure*}
In the case of SiMOS, fabrication begins on a natural silicon wafer, with 1 um of intrinsic silicon grown, followed by 100 nm epilayer ${}^{28}$Si, and a 10 nm thermally grown oxide \cite{Sabbagh2018}. First, tungsten (W) markers are patterned, which are used to define implant windows via electron beam lithography (EBL). After exposure, phosphorus ions (P${}^+$) at 6 keV are implanted to create highly negatively doped (n${}^{++}$) regions in each die. An activation anneal is conducted in a rapid thermal processor (RTP) at 1000 ${}^o$C for 30 seconds. A buffered hydro-fluoric (BHF) etch removes oxide in bond-pad areas, where Ti:Pt (5:55 nm) metallic contacts are deposited, creating ohmic contacts. A second layer of Ti:Pt markers are also written in this step. Next, a blanket Al${}_2$O${}_3$ ALD layer of 10 nm is grown across the entire sample. A small 20 x 20 $\mu$m${}^2$ area is exposed and etched away in the vicinity of the quantum dot formation area. This improves dot stability (see above). Large rounded rectangular regions are then exposed in regions where wirebonding is expected, and 150 nm of SiN is sputtered. These create safer bondpads with which to bond to, reducing leakage and improving device yield. The SiMOS device presented in the work utilizes a three layer Ti:Pd gate stack. (3:17, 3:37, 3:37 nm). After each layer, the device is annealed in a RTP furnace for 15 minutes at 400 ${}^o$C in forming gas, then a layer of ALD is grown at 7 nm thickness. Next, the qubit control layer is deposited. This can either be an Al or NbTiN antenna of 100 nm thickness for Electron spin resonance driving, or a Ti:Co micromagnet (5:195 nm) for Electron dipole spin resonance. The final step is an end of line anneal at 400 ${}^o$C for 30 minutes in forming gas in a RTP. \\
\newline

The Si/SiGe 5 dot linear array begins on a natural silicon substrate. A linearly graded Si${}_{1-x}$Ge${}_{x}$ layer is deposited where x ranges from 0 to 0.3. A relaxed Si${}_{0.7}$Ge${}_{0.3}$ layer of 300 nm lies below the 10 nm ${}^{28}$Si (800 ppm purity) quantum well which itself is separated from the 2 nm Si capping layer by a second 30 nm relaxed Si${}_{0.7}$Ge${}_{0.3}$  spacer layer. The initial marker layer is written using optical lithography and is formed by etching away the SiO${}_2$. Next, a BHF dip removes native oxide selectively where ohmic contacts of Ti:Pt (5:55 nm) are evaporated, alongside a second set of markers.  
Gate stack fabrication of the device is almost identical to that of SiMOS. It is a 3-layer Ti:Pd stack of the same thicknesses, interlayer isolated via 7 nm of Al${}_2$O${}_3$. However, we do not employ a gate anneal between gate layers, despite this technically being possible within the context of thermal budget. For control, both striplines and micromagnets are avaliable, however we prefer the electrical driving option since electron wavefunctions in Si/SiGe tend to be more mobile and hence EDSR provides a route to faster driving. We do not conduct an end of line anneal on SiGe devices. \\
\newline

For the fabrication of the Ge/SiGe 2x2 array, we begin with a natural silicon substrate, upon which 1.4 $\mu$m of Ge and 900 nm of reverse graded Si${}_{1-x}$Ge${}_{x}$ where x ranges from 1 to 0.8 is grown. This lies below a 160 nm  Si${}_{0.2}$Ge${}_{0.8}$ spacer layer, a 16 nm Ge quantum well under compressive strain, a second Si${}_{0.2}$Ge${}_{0.8}$ layer of 22 nm and finally a thin Si cap of 1 nm\cite{Sammak2019}. Ti:Pt EBL markers are then defined for future alignment. A short HF acid etch is conducted immidiately before depositing 30 nm Al on regions where ohmic contact is desired. An advantage of the Ge/SiGe platform is the possibility of ohmic formation extremely close (within $\approx100$ nm) of the quantum dot. Devices are then placed under vacuum for 1 h at 300 ${}^o$C causing Al to diffuse through the heterostructure into the quantum well forming ohmic contact. Atomic Layer Deposition is then performed covering the sample in a 10 nm Al${}_2$O${}_3$ blanket. The gate stack consists of two layers, barrier and plunger. The barrier layer is deposited at 20 nm total thickness utilizing the Ti:Pd stack (3:17 nm). The plunger layer is deposited at 40 nm total thickness (3:37 nm). No further processing is required as the large intrinsic Spin-Orbit coupling of holes in Ge/SiGe provides a native electric driving mechanism\cite{Hendrickx2019a}. 

\clearpage

\providecommand{\refin}[1]{\\ \textbf{Referenced in:} #1}

\end{document}